\documentclass[superscriptaddress,groupedaddress,nofootnoteinbib,12pt]{article} 
\pdfoutput=1
\usepackage{jcappub}
\usepackage{bm}
\usepackage{verbatim}

\usepackage{epsf}
\usepackage{graphicx,epsfig}
\usepackage{amsfonts}
\usepackage{amssymb}
\usepackage{xcolor}
\usepackage{textcomp}
\usepackage{subfigure}

\definecolor{mine}{rgb}{0.2,0.1,0.7}
\definecolor{bb}{rgb}{0.3, 0.5, 1}
\definecolor{bg}{rgb}{0.1, 0.1, 0.5}

\def\R{\zeta}

\def\a{\alpha}
\def\CL{{\cal L}}
\def\CH{{\cal H}}

\def\bkone{\mathbf k_1}
\def\bktwo{\mathbf k_2}
\def\bkthree{\mathbf k_3}
\def\bkfour{\mathbf k_4}

\def\p{\pi}

\def\nn{\nonumber}

\newcommand\bk{\boldsymbol{k}}

\newcommand{\bea}{\begin{eqnarray}}
\newcommand{\eea}{\end{eqnarray}}
\newcommand\be{\begin{equation}}
\newcommand\ee{\end{equation}}
\newcommand\beq{\begin{equation}}
\newcommand\eeq{\end{equation}}

\def\ba{\begin{eqnarray}}
\def\ea{\end{eqnarray}}

\newcommand{\refeq}[1]{(\ref{#1})}
\def\Tdot#1{{{#1}^{\hbox{.}}}}

\def\c{c_{{\cal D}}}

\def\ddc{{\ddot c}_{{\cal D}}}
\def\dc{{\dot c}_{{\cal D}}}
\def\inv{\frac{1}{\c^2}-1}
\def\g{g}
\def\tM{{\tilde M}}
\def\l{\lambda}
\def\M{M_{{\rm Pl}}}
\def\tg{{\tilde g}}

\begin{document}

\title{DBI Galileon in the Effective Field Theory of Inflation: Orthogonal non-Gaussianities and constraints from the Trispectrum}

\author[1,2]{S\'ebastien Renaux-Petel}
\affiliation[1]{Laboratoire de Physique Th\'eorique et Hautes Energies, University Paris 6, 4 place Jussieu, 75252 Paris, France.}
\affiliation[2]{Sorbonne Universit\'es, Institut Lagrange de Paris (ILP), 98 bis Bd. Arago, 75014 Paris, France.}
\emailAdd{srenaux@lpthe.jussieu.fr}
\vskip 4pt

\date{\today}


\abstract{Very few explicit inflationary scenarios are known to generate a large bispectrum of orthogonal shape. Dirac-Born-Infeld Galileon inflation, in which an induced gravity term is added to the DBI action, is one such model. We formulate it in the language of the effective field theory of inflation by identifying the unitary gauge operators that govern the behavior of its cosmological fluctuations. We show how to recover rather easily from this its power spectrum and bispectrum, which we calculated previously using standard cosmological perturbation theory. We push our calculations up to the determination of the fourth-order action and of the trispectrum, in which shapes absent in k-inflation arise due to the presence of higher-order derivative operators. We finally discuss the combined constraints set on this model by current observational bounds on the bispectrum and trispectrum.}


\maketitle


\section{Introduction}

The deviation from perfect Gaussian statistics of the primordial fluctuations promises to allow a more precise discrimination between competing scenarios of the early Universe than has hitherto been possible (see for instance \cite{Chen:2010xka,Langlois:2011jt} for recent reviews). In this respect, one of the important realization of the last decade has been the identification of a dictionary between, on one side, various mechanisms for generating the primordial curvature perturbation $\zeta$ and, on the other side, various momentum-dependences of its three-point correlation function, \textit{i.e.} various shapes of the primordial bispectrum \cite{Babich:2004gb,Fergusson:2008ra}. The three best known shapes, local \cite{Komatsu:2001rj}, equilateral \cite{Creminelli:2005hu} and orthogonal \cite{Senatore:2009gt}, have two distinct well understood physical origins. The former is generated classically by the non-linear evolution of the primordial fluctuations on super-Hubble scales and it is associated to the presence of several degrees of freedom during inflation \cite{Maldacena:2002vr,Creminelli:2004yq} (or other competing early-universe scenarios). The last two are generated quantum-mechanically by derivative interactions that force correlations amongst fluctuations of similar wavelengths, in models with non-standard kinetic terms, single- or multi-field \cite{Alishahiha:2004eh,Chen:2006nt,Langlois:2008wt,Langlois:2008qf}. The difference between the equilateral and the orthogonal shape then only results from details of the amplitudes at which various derivative interactions get excited, which is naturally formulated in the language of the effective field theory of inflation \cite{Cheung:2007st}. 

In terms of explicit scalar field realizations, archetypal scenarios that generate the local and the equilateral shapes are well known, like the curvaton \cite{Lyth:2001nq} and DBI inflation \cite{Silverstein:2003hf,Alishahiha:2004eh} respectively\footnote{Naturally, these two shapes do not constitute separate possibilities and a linear combination of them can be present in the same model, like in multifield DBI inflation \cite{RenauxPetel:2009sj}.}. Similarly, we find it important to have at hand concrete microphysical early-universe scenarios able to generate a large orthogonal bispectrum, all the more so as the WMAP nine-year analysis indicated a (non-significant) $2.45\sigma$ hint of such a signal \cite{Bennett:2012fp}. Dirac-Born-Infeld Galileon (DBI Galileon) inflation, considered in references \cite{RenauxPetel:2011dv,RenauxPetel:2011uk}, provides the first and one of the very few examples of such a model\footnote{Interestingly, a bispectrum with a significant overlap with the orthogonal shape was recently shown to arise naturally in a different setup \cite{Green:2013rd}.}. Its origin is physically transparent and its action can be written in simple geometrical terms: it consists in supplementing the DBI action, describing the motion of a probe D3-brane, with an induced gravity term. It can hence effectively be seen as an extension of the well known DBI inflationary scenario with only one new dimensionless parameter setting the strength of the induced gravity. When varying the latter, the shape of the bispectrum varies continuously, interpolating between highly correlated and highly anti-correlated with the equilateral template, while it is of the orthogonal type in a transitionary region.\\

The purpose of the present paper is two-fold. As the orthogonal shape has been first motivated and constructed in the effective field theory formalism, we would like to formulate the DBI Galileon model in this language, \textit{i.e.} to identify the operators and related mass scales entering into the Lagrangian describing its fluctuations. As we will see, the way the orthogonal shape appears in this model is similar to, but different in details from, the one in its original construction in \cite{Senatore:2009gt}. Our second aim is to investigate the possibility of observationally falsifying DBI Galileon inflation as an inflationary scenario candidate for explaining the large orthogonal bispectrum suggested in WMAP data \cite{Bennett:2012fp}. Indeed, we recently pointed out the fact that the trispectrum can be used, already with current data, to constrain such candidates in the simplest theoretical framework \cite{Renaux-Petel:2013wya}. Although the two setups are different, we will see that similar conclusions can also be reached here, and that the trispectrum is putting the DBI Galileon model under strong observational pressure, at least in the $1\sigma$ region favored by WMAP9 constraints on the bispectrum.\\

The plan of our paper is the following: in section \ref{sec:background}, we introduce the DBI Galileon model and briefly discuss its background evolution. In section \ref{sec:EFT}, we formulate it in the language of the effective field theory of inflation. We show in particular how to use the latter to recover rather easily the leading-order quadratic and cubic actions in the cosmological fluctuations, and hence the power spectrum an bispectrum determined in \cite{RenauxPetel:2011dv,RenauxPetel:2011uk}. We also derive the action quadratic in fluctuations, and the primordial trispectrum generated in this model. The section \ref{constraints} is then devoted to a discussion of its observational status. We conclude in section \ref{sec:conclusion} and collect technical details, lengthy results and some plots in three appendices.

\section{DBI Galileon: action and background}
\label{sec:background}

In this section, we briefly review the aspects of the DBI Galileon model that are relevant for what follows, referring the reader to \cite{RenauxPetel:2011dv,RenauxPetel:2011uk} for an extensive study of it, in particular its multifield aspects. For simplicity, we consider only its single-field version here, or equivalently we neglect the influence of the entropic perturbations on the adiabatic one. The action we consider reads
\be
S= \int {\rm d}^4 x \left[ \frac{\M^2}{2}  \sqrt{-g} R[g]  + \sqrt{-g}\,  {\cal L}_{\rm DBI} +\frac{\tM^2}{2} \sqrt{-\tg} R[\tg]    \right] \,, 
\label{action-f=cst}
\ee
where the first term is the standard Einstein-Hilbert action involving the cosmological metric $g_{\mu \nu}$, the second term is the Dirac-Born-Infeld action \cite{Silverstein:2003hf}
\be
 {\cal L}_{\rm DBI}= -\frac{1}{f}\left(\sqrt{1+f g^{\mu \nu} \partial_{\mu} \phi   \partial_{\nu} \phi }-1\right) -V(\phi)\,,
 \label{brane-action}
\ee
and the third term is proportional to the Einstein-Hilbert Lagrangian of the so-called induced metric\footnote{In the higher-dimensional framework which gives rises to the effective four-dimensional Lagrangian \refeq{action-f=cst}, the metric \refeq{induced-tg} is proportional to the induced metric on the brane, hence its name.}
\be
\tg_{\mu \nu} \equiv  g_{\mu \nu}  + f \, \partial_\mu \phi \partial_\nu \phi\,.
\label{induced-tg}
\ee
We also consider a constant warp-factor $f$ for simplicity (see \cite{RenauxPetel:2011uk} for a discussion of the more general case), acknowledging that the effects of a non-constant $f$ would anyway be negligible in a slow-varying approximation.

In a spatially flat Friedmann-Lema\^itre-Robertson-Walker (FLRW) background spacetime, of metric
\be
g_{\mu \nu} dx^{\mu} dx^{\nu}=-dt^2 +a^2(t) d \boldsymbol{x}^2\,,
\ee
 the background equations of motion derived from the action \refeq{action-f=cst} can be cast in the simple form\footnote{As usual in single-field inflation, the equation of motion derived from varying the action \refeq{action-f=cst} with respect to $\phi$ is a consequence of the gravitational equations \refeq{Friedmann1}-\refeq{Hdot}.} \cite{RenauxPetel:2011dv,RenauxPetel:2011uk}
 \bea
\label{Friedmann1}
 3 H^2 \left(\M^2 +\frac{\tM^2}{\c^3} \right)= V+ \frac{1}{f}\left( \frac{1}{\c}-1 \right)
\eea
\bea
\M^2 H^2 \epsilon +\frac{ \tM^2 H^2}{\c} \left( \epsilon +s+\frac32 \left( \inv \right) \right)= \frac{{\dot \phi}^2}{2\c}\,,
\label{Hdot}
\eea
where $\epsilon \equiv -{\dot H}/H^2$ denotes the standard inflationary parameter, with $H \equiv {\dot a}/a$ being the Hubble parameter,
\be
\c^2 \equiv 1- f  \dot \phi^2\,,
\ee
and where $s \equiv {\dot c_{{\cal D}}}/(H \c)$\footnote{We use the symbol $\c^2$, and not the more conventional $c_s^2$, because, contrary to standard brane inflation, $\c$ does not coincide here with the speed of sound of scalar perturbations.}. Like in standard brane inflation, the value of $\c^2$ distinguishes the slow-roll regime, where $\c^2 \simeq 1$ and the brane action \refeq{brane-action} takes a canonical form, from the so-called relativistic, or DBI, regime, where the inflaton almost saturates its speed limit (${\dot \phi}^2 < 1/f$) and $\c^2  \ll 1$. It is this regime, where the non-linear structure of  the action \refeq{brane-action} must be taken into account and large non-Gaussianities are generated, that we are particularly interested in. From Eqs.~\refeq{Friedmann1}-\refeq{Hdot}, one can show that achieving a phase of quasi de-Sitter expansion -- $\epsilon \ll 1$ -- in the relativistic regime, requires, in addition to the usual condition to enter into the DBI regime 
\be
\c f V \gg 1
\label{condition1}\,,
\ee
that 
\be
\tM^2 \ll \c^3 \M^2\,.
\label{condition2}
\ee
Despite this restrictive condition on the mass scale associated to the induced gravity, the effects of the latter on the cosmological fluctuations can actually be significant, and are conveniently measured by the dimensionless parameter
\be
\alpha \equiv \frac{ f H^2 \tM^2}{\c^2}\,.
\label{def-alpha}
\ee
The latter is not necessarily small during inflation, although we will see it has to be less than $1/9$ to avoid the presence of ghosts. In the inflationary regime of interest where the conditions \refeq{condition1} and \refeq{condition2} are fulfilled, note eventually that Eq.~\refeq{Hdot} reduces to 
\be
\epsilon \simeq  \frac{1}{2 \c f  H^2 \M^2}  (1-3\alpha) \,.
\label{eps-approx} 
\ee
At leading order in a slow-varying approximation, we can consider $\c$ and $\alpha$ as constant, and we will see in section \ref{constraints} how observational constraints on the primordial bispectrum and trispectrum directly translate into constraints on these two parameters.

\section{DBI Galileon: fluctuations in the effective field theory}
\label{sec:EFT}

In references \cite{RenauxPetel:2011dv,RenauxPetel:2011uk}, we calculated the second- and third-order action of cosmological fluctuations in DBI Galileon inflation, from which we deduced the corresponding power spectrum and bispectrum of the curvature perturbation $\zeta$. In this section, we show how these results can be recovered in an elegant way in the effective field theory formalism. Moreover, the latter is powerful enough that we are able to derive rather easily the fourth-order action, as necessary for the study of the primordial trispectrum (see \cite{Senatore:2010jy,Bartolo:2010di} for other applications of the effective field theory of inflation to the study of the primordial trispectrum).

We refer the reader to \cite{Creminelli:2006xe,Cheung:2007st} for details about the construction of the effective field theory of inflation, only summarizing here the strategy we followed in our calculation: we first expressed the action \refeq{action-f=cst} in the unitary gauge in which the scalar field $\phi$ is unperturbed (subsection \ref{subsec:unitary}), paying a particular attention to put our result in a convenient form for the second part of the calculation (subsection \ref{subsec:pi}), where we used the St\"uckelberg trick to explicitly reintroduce the scalar perturbation $\pi$ and derive the action governing its dynamics in the decoupling regime in which the mixing with gravity is irrelevant.

\subsection{Action in the unitary gauge}
\label{subsec:unitary}

We set ourselves in the unitary gauge in which the scalar field is unperturbed, \textit{i.e.} such that $\phi=\phi(t)$, and use the ADM formalism \cite{adm} in which the cosmological metric is written in the form
\beq
g_{\mu \nu} d x^{\mu} dx^{\nu}=-N^2 dt^2 +h_{ij} (dx^i+N^i dt)(dx^j+N^j dt)\,,
\label{g-ADM}
\eeq
where $h_{ij}$ is the spatial metric, $N^i$ is the shift vector and $N$ the lapse function. In the unitary gauge, the induced metric \refeq{induced-tg} takes the very simple form
\beq
\tg_{\mu \nu} dx^{\mu} dx^{\nu}=- {\tilde N}^2  dt^2 +h_{ij} (dx^i+N^i dt)(dx^j+N^j dt)
\eeq
where only the lapse function
\be
{\tilde N}^2 \equiv N^2-1+\c^2
\ee
is modified compared to Eq.~\refeq{g-ADM}. Using the ADM form of the Einstein-Hilbert action, the induced gravity term $S_{\rm ind}= \frac{\tM^2}{2}  \int {\rm d}^4 x \sqrt{-\tg} R[\tg]$ reads

\be
S_{\rm ind}= \frac{\tM^2}{2}   \int {\rm d}^4 x \sqrt{-g} \left( \frac{\tilde N}{N}R^{(3)} + \frac{N}{\tilde N}\left( K_{ij} K^{ij}-K^2\right) \right)
\label{Sind}
\ee
where $R^{(3)}$ is the scalar Ricci curvature of the spatial metric $h_{ij}$ and $K_{ij} \equiv \frac{1}{2N} \left({\dot h}_{ij}-2 \nabla_{(i} N_{j)} \right)$ is the extrinsic curvature of constant time hypersurfaces. In the effective field theory of inflation as formulated in \cite{Cheung:2007st}, the $3$d scalar Ricci curvature $R^{(3)}$ is traded for the four-dimensional one $R$ using Gauss-Codazzi relations. While this is useful for classification purposes in a general formalism, it turns out here to be easier to use the two variables for practical calculations, as also noted in \cite{Bloomfield:2012ff}. Using \cite{Cheung:2007st}
\be
R^{(3)}=R+K^2- K_{ij} K^{ij}-2\nabla_{\mu}(n^{\mu} \nabla_{\nu} n^{\nu})+2\nabla_{\nu}(n^{\mu} \nabla_{\mu} n^{\nu})\,,
\ee
where $n^{\mu}$ is the unit vector perpendicular to constant time hypersurfaces (we choose it such that $n^0 >0$), and the fact that $g^{00}=-1/N^2$, we therefore conveniently write
\bea
 \int {\rm d}^4 x \sqrt{-g} \frac{\tilde N}{N}R^{(3)} &=& \int {\rm d}^4 x \sqrt{-g}  \left(\sqrt{1+(1-\c^2)g^{00}}-\c \right) R^{(3)} \nn \\
 &+& \int {\rm d}^4 x \sqrt{-g}\, \c \left(R+K^2- K_{ij} K^{ij}-2\nabla_{\mu}(n^{\mu} \nabla_{\nu} n^{\nu}) \right)
 \label{R3}
\eea
where in the last line we used the fact that \cite{Cheung:2007st}
\be
 \int {\rm d}^4 x \sqrt{-g} \, \c(t) \nabla_{\nu}(n^{\mu} \nabla_{\mu} n^{\nu})=- \int {\rm d}^4 x \sqrt{-g}\, \partial_{\nu}(\c(t))\, n^{\mu} \nabla_{\mu} n^{\nu}=0
\ee
as $\partial_{\nu}(\c(t)) \propto n_{\nu}$. While the advantage of the mixed form Eq.~\refeq{R3} may not be obvious at this stage, it will become transparent in the next subsection where we will explicitly reintroduce the scalar fluctuation $\pi$. Using Eqs.~\refeq{Sind} and \refeq{R3}, we can finally rewrite the full action \refeq{action-f=cst} in the form (convenient for what follows)
\bea
S&=& \int {\rm d}^4 x \sqrt{-g}  \frac{\M^2}{2} \left(1+\c \frac{\tM^2}{\M^2} \right)R\nn \\
&+&  \frac{\tM^2}{2} \int {\rm d}^4 x \sqrt{-g}  \left( \frac{1}{\sqrt{1+(1-\c^2)g^{00}}} -\c \right) \left( K_{ij} K^{ij}-K^2\right)  \nn \\
&+&  \frac{\tM^2}{2}  \int {\rm d}^4 x \sqrt{-g}  \left(\sqrt{1+(1-\c^2)g^{00}}-\c \right) R^{(3)} \nn \\
&+& \tM^2  \int {\rm d}^4 x \sqrt{-g}\, n^{\mu}  \partial_{\mu}( \c(t))\, \nabla_{\nu} n^{\nu} \nn \\
&+& \int {\rm d}^4 x \sqrt{-g} \left(   -\frac{1}{f}\left(\sqrt{1+(1-\c^2)g^{00} }-1\right) -V(t) \right)\,.
\label{S-unitary}
\eea

\subsection{Reintroducing the Goldstone boson}
\label{subsec:pi}

We aim at deriving the action governing the dynamics of the cosmological fluctuations about the background discussed in section \ref{sec:background}. For this purpose, and following \cite{Cheung:2007st,Gubitosi:2012hu,Bloomfield:2012ff}, we separate Eq.~\refeq{S-unitary} into
\bea
S=\int d^4x \sqrt{-g} \left[ \frac{\M^2}{2} \g(t) R-\Lambda(t)-c(t)g^{00} \right]+S^{(n\geq 2)}
\label{S-EFT}
\eea
where $S^{(n\geq 2)}$ explicitly starts quadratic in the fluctuations $\delta g^{00} \equiv g^{00}+1$, $\delta K_{ij}=K_{ij}- H h_{ij}$ and $\delta R^{(3)}=R^{(3)}$. As explained in \cite{Cheung:2007st}, $g(t)$ can be set to one by a conformal transformation of $g_{\mu \nu}$. While we could follow this strategy here, we find it more convenient to keep a non-trivial $g(t)$, as arising naturally from the first line in Eq.~\refeq{S-unitary}. In this sense, although we are dealing with a phase of single-clock inflation, we are actually using the tools of the effective field theory of dark energy as developed in \cite{Gubitosi:2012hu,Bloomfield:2012ff} (note that the difference between the formalisms of the effective field theory of inflation and of dark energy is mainly semantic, the two being primarily effective field theories of adiabatic fluctuations about a FLRW background; see the seminal reference \cite{Creminelli:2006xe} where applications to both contexts were first considered). We leave the explicit form of $g(t)$, $\Lambda(t)$ and $c(t)$ to the appendix \ref{Appendix:calculations}, where we also give details of the intermediate steps that lead to the result $S^{(n\geq 2)}=\int d^4x \sqrt{-g} {\cal L}^{(n\geq 2)}$ with
\bea
{\cal L}^{(n\geq 2)}&=& \frac{(1-9\a)}{8 f \c^3}(\delta g^{00})^2 +\frac{H \tM^2}{\c^3}  \delta g^{00} \delta K+\frac{\tM^2}{2 \c}(\delta K_{ij} \delta K^{ij}-\delta K^2) +\frac{\tM^2}{4 \c}\delta g^{00} \delta R^{(3)} \label{Main}  \\
 && \hspace{-4.0em} -  \frac{(1-15 \a)}{16 f \c^5} (\delta g^{00})^3-\frac{3  H\tM^2}{4\c^5} (\delta g^{00})^2  \delta K -\frac{\tM^2}{4 \c^3}\delta g^{00} (\delta K_{ij} \delta K^{ij}-\delta K^2) -\frac{\tM^2}{16 \c^3} (\delta g^{00})^2 \delta R^{(3)} \nn \\
&& \hspace{-4.0em} + \frac{5(1-21 \a)}{128 f \c^7} (\delta g^{00})^4+\frac{5  H\tM^2}{8\c^7} (\delta g^{00})^3  \delta K+\frac{3 \tM^2}{16 \c^5} (\delta g^{00})^2 (\delta K_{ij} \delta K^{ij}-\delta K^2) +\frac{\tM^2}{32 \c^5} (\delta g^{00})^3 \delta R^{(3)} \nn
\eea
up to quartic order in fluctuations. Here, we kept leading order terms in the relativistic regime $\c^2 \ll 1$ and in a slow-varying approximation. In particular, we neglected terms involving time-derivatives of $\c$ (it is easy to take them into account if needed), arising for instance from the fourth line in Eq.~\refeq{S-unitary}. This is an important result of our paper: the identification of the operators and related mass scales that specify the DBI Galileon model in the effective field theory of inflation in the unitary gauge. The first line (respectively second and third) contains the terms that start explicitly quadratic (respectively cubic and quartic) in the fluctuations. Note that the result simplifies considerably for the DBI model, corresponding to $\tM^2=\alpha=0$, and for which the only present operators are powers of $\delta g^{00}$.

 While Eq.~\refeq{Main} is an important step in our calculation, the true effectiveness of the effective field theory of inflation, however, relies on the gravitational analogue of the equivalence theorem for the longitudinal components of a massive gauge boson \cite{Cornwall:1974km}: upon performing a time diffeomorphism $t \to t+\pi(x)$ on the gauge-fixed action \refeq{S-unitary}-\refeq{S-EFT}, thus restoring full time-diffeomorphism invariance\footnote{The field $\pi$ is assumed to transform under the time diffeomorphism ${\tilde t}=t+\xi^0(x^{\mu})$ as ${\tilde \pi} =\pi-\xi^0(x^{\mu})$, such that $t+\pi$ is invariant.}, one can show that the scalar fluctuation $\pi$ reintroduced this way decouples from the gravitational sector at high enough energies, allowing to neglect the complications of the mixing with gravity. This latter fact was also shown to be valid in the relativistic regime of DBI Galileon inflation in \cite{RenauxPetel:2011dv,RenauxPetel:2011uk}, and we consider only this decoupling regime in what follows.

 The effect of the St\"uckelberg diffeomorphism $t \to t+\pi(x)$ on explicit functions of time is $h(t) \to h(t+\p) \approx h(t)+\dot h(t) \p+\frac{\ddot h(t)}{2} \p^2+\ldots$ The terms in $\pi$ generated in this process are negligible in the slow-varying approximation and we don't keep track of them, consistently with the fact that we discarded terms involving derivatives of $\c$. Restoring $\pi$ thus only affects the operators through the replacements \cite{Cheung:2007st,Gubitosi:2012hu,Bloomfield:2012ff}
 \bea
\delta g^{00} & \to & -2 {\dot \pi}-{\dot \pi}^2+\frac{(\partial_i \pi)^2}{a^2}\,, \label{replace-1}  \\
\delta K_{ij} &\to & -{\dot H} \pi h_{ij}-\partial_{i}\partial_{j} \pi\,, \\
\delta K &\to & -3 {\dot H} \pi-\frac{\partial^2 \pi}{a^2}\,, \\
\delta R^{(3)} &\to & 4H \frac{\partial^2 \pi}{a^2} \,, \label{replace-4}
\eea
again, in the decoupling regime in which the metric is left unperturbed. Note eventually that $\pi$ is related to the curvature perturbation $\zeta$ by the simple relation $\zeta=-H \pi$  at linear order and at leading order in a slow-varying approximation \cite{Cheung:2007st}.

\subsubsection{Quadratic action and power spectrum}

From the above formulae, it is straightforward to derive the main second-order action
\bea
S^{(2)}&=&\int dt \, d^3 x \, a^3\, \left[    \frac{(1-9 \a)}{2 f \c^3} \dot \p^2   -\frac{(1-5 \a)}{2 f \c} \frac{(\partial_i \p)^2}{a^2} \right]\,,
\label{S2-1}
\eea
in agreement with the one found in \cite{RenauxPetel:2011dv,RenauxPetel:2011uk}, and where we used the integration by part $\int dt \, d^3 x \, a^3\, {\dot \pi} \frac{ \partial^2 \pi}{a^2}=\int dt \, d^3 x \, a\, \frac{H}{2} (\partial_i \pi)^2$. As mentioned in section \ref{sec:background}, $\pi$ becomes a ghost when $\alpha > 1/9$, hence our limitation to the range $0\leq  \alpha < 1/9$. From Eq.~\refeq{S2-1}, one deduces the speed of sound $c_s$ at which the curvature perturbation propagates:
\be
c_s \equiv \c \sqrt{\frac{1-5\alpha}{1-9 \alpha}}\,.
\label{cs}
\ee
Note that for any $\c$, there is a limiting value of $\alpha$ above which $c_s$ in Eq.~\refeq{cs} is greater than one. However, in the relativistic regime in which our calculations are applicable, for $\c \lesssim 0.1$, these values are very close to $1/9$. We will see that the theory becomes strongly coupled as $\alpha \to 1/9$, not surprisingly as the kinetic term in \refeq{S2-1} goes to zero in this limit. For practical purposes, there is thus no issue of superluminality, as the region where it arises is already excluded by observational bounds on non-Gaussianities.

In the almost de-Sitter regime of interest, in which $a\simeq -1/(H \tau)$ with $\tau$ the conformal time such that $dt=a\, d \tau$, the linear equation of motion derived from Eq.~\refeq{S2-1} is simply
\be
{\ddot \pi}+3H {\dot \pi}-c_s^2 \frac{\partial^2 \pi}{a^2}=0\,,
\label{linear-eom}
\ee
whose standard solution with Bunch-Davies normalization and initial conditions reads (in Fourier space)
\be
\pi_k(\tau)=\sqrt{\frac{1-3 \alpha}{1-5\alpha}}\frac{1}{\sqrt{4 \epsilon c_s k^3} \M}(1+i k c_s \tau) e^{-i k c_s \tau}\,.
\label{standard-zeta}
\ee
We then easily deduce the power spectrum of the curvature perturbation:
\be
{\cal P}_{\zeta} \equiv \frac{k^3}{2 \pi^2} |\zeta_k(\tau \to 0) |^2 = \frac{1-3 \alpha}{1-5\alpha} \frac{1}{8 \pi^2 \epsilon c_s} \frac{H^2}{\M^2}\,.
\label{P-zeta}
\ee

\subsubsection{Cubic action and bispectrum}
\label{subsec:3}

\begin{figure}[!h]
  \center
  \includegraphics[width=0.7\textwidth]{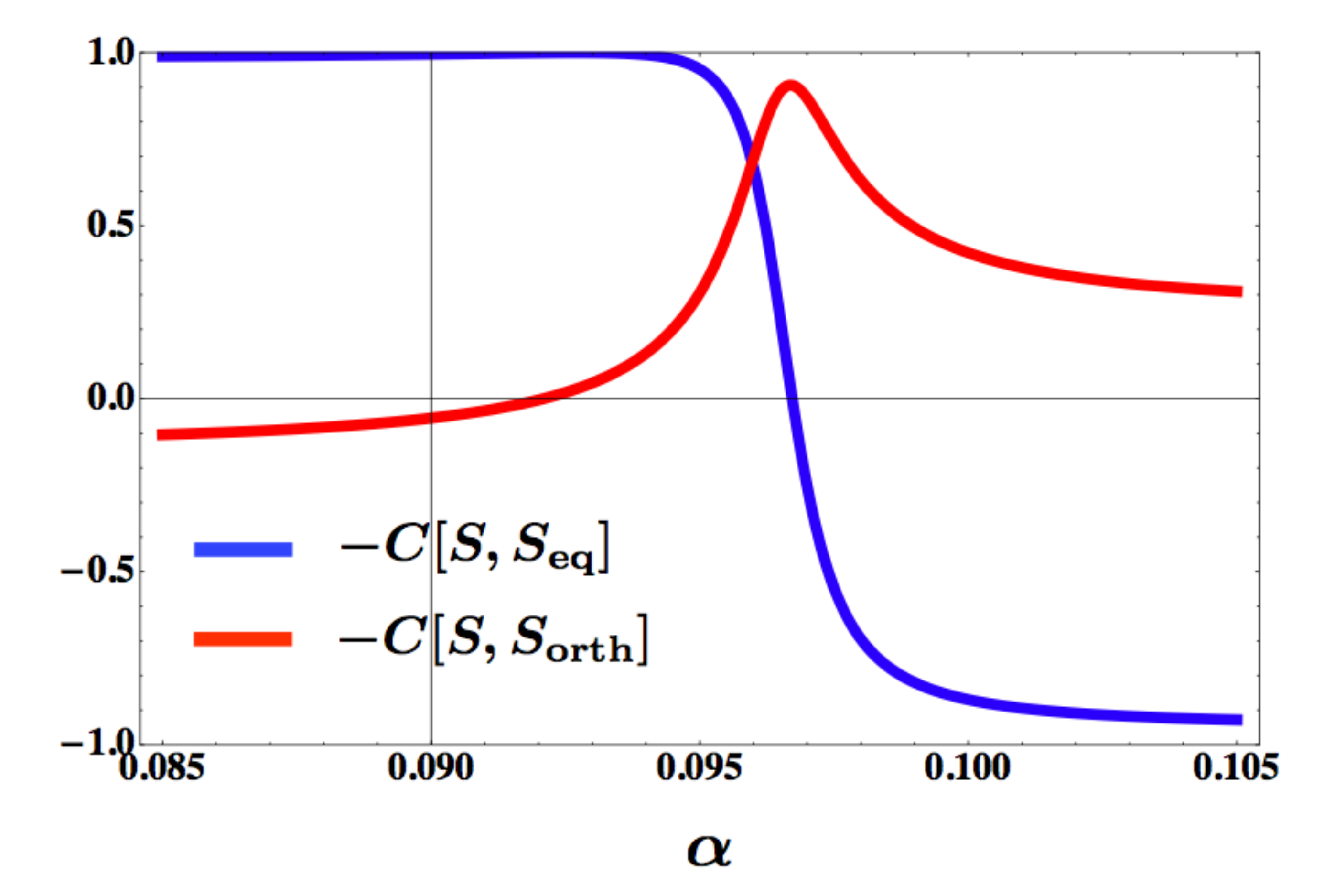}
      \caption{Correlation of (minus) the DBI Galileon shape \refeq{S} with the equilateral template (blue) and the orthogonal template (red) as a function of $\alpha$.}
      \label{fig:Bispectrum-correlations}
\end{figure}

Around the time of Hubble crossing $k c_s \tau \approx -1$, it is straightforward to verify from Eq.~\refeq{standard-zeta} that $\dot \p \sim H \p$ and $\partial_i \p/a \sim H \p/\c$. Taking this into account (note in particular that amongst terms with the same number of derivatives, the ones with the highest number of space derivatives dominate in the relativistic regime $\c^2 \ll 1$), we deduce from Eqs.~\refeq{Main} and \refeq{replace-1}-\refeq{replace-4} the leading-order cubic action:
\bea
S^{(3)}&=&\int dt \, d^3 x \, a^3\, \left[    \frac{(1-15 \a)}{2 f \c^5} \dot \p^3   -\frac{(1-9 \a)}{2 f \c^3} \dot \p \frac{(\partial_i \p)^2}{a^2} 
\right.
 \cr
  &&  \hspace{6.0em} 
  \left.
+\frac{3 H \tM^2}{\c^5} \dot \p^2 \frac{\partial^2 \p}{a^2}  -\frac{3H \tM^2}{4\c^3} \frac{(\partial_i \p)^2 \partial_j^2 \p}{a^4} \right]
 \label{S3-1}
\eea
where we made use of 
\be
\int dt \, d^3 x \, \frac{{\dot \pi}}{a} \left((\partial_i \partial_j \p)^2-(\partial_i^2 \p)^2 \right)=\int dt \, d^3 x \, \frac{H}{2 a} (\partial_i \p)^2 \partial_j^2 \p\,,
\ee
derived by successive simple integrations by part. Note that the cubic action \refeq{S3-1} contains terms involving second-order (spatial) derivatives of $\pi$ (in its second line) as a result of the presence of extrinsic curvature terms in the unitary gauge action \refeq{Main}. However, as it was explained in a general context in references \cite{Arroja:2011yj,Burrage:2011hd,RenauxPetel:2011sb}, it is legitimate to simplify the interacting action of cosmological fluctuations by using the linear equation of motion derived from the quadratic action. Using this fact, one can actually trade the two higher derivative operators $ \dot \p^2 \partial^2 \p$ and $(\partial_i \p)^2 \partial_j^2 \p$ for the single-derivative ones ${\dot \pi}^3$ and ${\dot \pi} (\partial_i \pi)^2$ (see for instance \cite{Creminelli:2010qf} and Eqs. $4.5$ and $4.6$ in \cite{RenauxPetel:2011uk}). The effective third-order action then reads
\bea
 \hspace{-2.0em}  S^{(3)}_{\rm eff}&=&\int dt \, d^3 x \, a^3\,  \frac{1}{2 f \c^5}   \left[A_{{\dot \pi}^3}  \dot \p^3 - A_{{\dot \pi} (\partial_i \p)^2} \c^2  \dot \p \frac{(\partial_i \p)^2}{a^2}    \right]
 \label{S3-2}
\eea
with
\bea
A_{{\dot \pi}^3}&=&1-3 \a(5-4 \l^2+\l^4) \label{A2}\,, \\
A_{{\dot \pi} (\partial_i \p)^2}&=& 1-\a(9-3 \l^2) \label{A1}\,, \\
\lambda &=&\frac{\c}{c_s}=\sqrt{\frac{1-9 \alpha}{1-5 \alpha}} \label{lambda}\,.
\eea
We thus recovered the cubic action derived in \cite{RenauxPetel:2011dv,RenauxPetel:2011uk}, albeit with considerably less effort thanks to the effective field theory formalism. The primordial bispectrum deduced from Eq.~\refeq{S3-2} reads  \cite{RenauxPetel:2011dv,RenauxPetel:2011uk}
\be
\langle \R(\bk_1) \R(\bk_2) \R(\bk_3) \rangle = (2 \pi)^7 \delta (\sum_{i=1}^3 \bk_i) {\cal P}_{\R}^2 \frac{S(k_1,k_2,k_3)}{(k_1 k_2 k_3)^2}
\label{S-def} 
\ee
where ${\cal P}_{\R}$ is the primordial power spectrum given in Eq.~\refeq{P-zeta} and the shape function 
\bea
S(k_1,k_2,k_3)=-\frac{1}{2\c^2} \frac{1}{ 1-9 \alpha} \left( 3 A_{{\dot \pi}^3}S_{{\dot \pi}^3}+ \frac{\l^2}{2}A_{{\dot \pi} (\partial_i \p)^2}  S_{{\dot \pi} (\partial_i \p)^2} \right)
\label{S}
\eea
is a linear combination (function of $\alpha$) of
\bea
S_{{\dot \pi}^3}&=&\frac{k_1 k_2 k_3}{K^3}
\label{S1}
\\
S_{{\dot \pi} (\partial_i \p)^2} &=& -\frac{k_3}{k_1 k_2 K^3}  \bk_1 \cdot \bk_2 \left(2 k_1 k_2 -k_3 K+   2K^2 \right)+ 2\,{\rm perms.} \
\label{S-other}
\eea
where $K \equiv  k_1+k_2+k_3 $. The bispectrum generated in DBI Galileon inflation can not be mapped to the one through which the orthogonal shape was introduced in reference \cite{Senatore:2009gt} (note though that the effects of higher-derivative operators are discussed there as well). Yet, its qualitative features are similar: its overall amplitude is fixed by the speed of sound $c_s$ (or equivalently by $\c$) while its shape depends continuously on a dimensionless parameter (here $\alpha$). In particular, the former interpolates between highly anti-correlated (respectively correlated) with the equilateral template as $\alpha \to 0$ (respectively $\alpha \to 1/9$), while being of the orthogonal type in a transitionary region centered around $\alpha=0.097$. This is shown in Fig.~\ref{fig:Bispectrum-correlations} where we plot the scalar product, in the sense defined in \cite{Fergusson:2008ra}, between the DBI Galileon shape \refeq{S} and the equilateral and orthogonal shapes as a function of $\alpha$ (see \cite{RenauxPetel:2011uk} for more details and section \ref{constraints} for observational constraints).

\subsubsection{Quartic action and trispectrum}
\label{subsec:4}

Following the same logic that led to the cubic action \refeq{S3-1} from Eqs.~\refeq{Main} and \refeq{replace-1}-\refeq{replace-4}, we find the leading-order quartic action
\bea
\hspace{-4.0em}  S^{(4)}&=&\int dt \, d^3 x \, a^3\, \left[    \frac{5(1-21 \a)}{8 f \c^7} \dot \p^4  -\frac{3 (1-15 \alpha)}{4 f \c^5} \dot \p^2 \frac{(\partial_i \pi)^2}{a^2}
 +\frac{(1-9 \a)}{8 f \c^3} \frac{((\partial_i \p)^2)^2}{a^4}  
\right.
 \cr
  &&\hspace{-3.5em}  + 
   \left.
\frac{\tM^2}{4\c^5} \left( 3{\dot \pi}^2-\c^2 \frac{(\partial_i \pi)^2}{a^2}  \right)   \frac{(\partial_i \partial_j \p)^2-(\partial_i^2 \p)^2}{a^4}
+
\frac{5 H \tM^2}{\c^7} \dot \p^3 \frac{\partial_i^2 \p}{a^2}-\frac{3 H \tM^2}{\c^5} \dot \p \frac{(\partial_i \p)^2 \partial_j^2 \p}{a^4}
\right].
\label{S4-1}
\eea
Similarly to the case of the cubic action, one can simplify \refeq{S4-1} by using the linear equation of motion \refeq{linear-eom}, leading to (see appendix \ref{Appendix:trispectrum} for intermediate steps):
\bea
\hspace{-1.0em}
  S^{(4)}_{{\rm eff}}&=&\int dt \, d^3 x \, a^3\,\frac{1}{8 f \c^7} \left[   5\left(1- \a \left(21-18 \l^2+\frac{18}{5} \l^4  \right)\right)\dot \p^4    
+  (1-9 \a) \c^4 \frac{((\partial_i \p)^2)^2}{a^4} 
\right.
 \cr
  &&\hspace{-3.0em}  
 -
     \left. 6 \left(1-5 \alpha \left(3-2 \l^2 \right)\right) \c^2 \dot \p^2 \frac{(\partial_i \pi)^2}{a^2}
  + 
\frac{2 \alpha \c^4}{H^2} \left( 3{\dot \pi}^2-\c^2 \frac{(\partial_i \pi)^2}{a^2}  \right)   \frac{(\partial_i \partial_j \p)^2-(\partial_i^2 \p)^2}{a^4}
\right]
\label{S4-2}
\eea
which can not be further simplified, \textit{i.e.} where the quartic operators are independent of each other. This fourth-order action can be usefully compared to its counterpart in k-inflation \cite{Chen:2009bc} (see also \cite{Arroja:2009pd}). One observes that the two operators in ${\dot \pi}^2((\partial_i \partial_j \p)^2-(\partial_i^2 \p)^2)$ and $(\partial_k \pi)^2((\partial_i \partial_j \p)^2-(\partial_i^2 \p)^2)$, which, as we stress again, can not be traded for the three other ones, are not generated in k-inflation. This is to be contrasted with the cubic action \refeq{S3-2} and is a direct consequence of the presence of higher-derivative operators involving extrinsic curvature terms in the unitary gauge action \refeq{Main}.

We find then that the primordial trispectrum generated in DBI Galileon inflation reads
\be
\langle \R(\bk_1) \R(\bk_2) \R(\bk_3)  \R(\bk_4) \rangle_c = (2 \pi)^9 \delta (\sum_{i=1}^4 \bk_i) {\cal P}_{\R}^3 \frac{{\cal T}(k_1,k_2,k_3,k_4,k_{12},k_{14})}{(k_1 k_2 k_3 k_4)^3}
\label{T-def} 
\ee
with $k_{ij}=   | \bk_{i}+\bk_{j}  |$ and
\bea
{\cal T}= \frac{1}{\c^4} \left( \sum_{i=1}^{5} A_{ci} T_{ci}+\sum_{j=1}^{3} A_{sj} T_{sj} \right)
\label{result-T}
\eea
being the sum of 8 contributions: 5 contact-interaction trispectra $T_{ci}$ -- corresponding to the 5 quartic operators in \refeq{S4-2} -- and 3 scalar-exchange trispectra $T_{sj}$, solely determined by the cubic (and quadratic) action  \refeq{S3-2}. We use the same notations as in \cite{Chen:2009bc} for $T_{c1}, T_{c2}, T_{c3}, T_{s1}, T_{s2}$ and $T_{s3}$, whose explicit and lengthy expressions can be found there, while we defined
\bea
T_{c4}&=&-k_1^2 k_2^2 ( (\bkthree \cdot \bkfour)^2-k_3^2 k_4^2) \times \nn \\
& & \frac{1}{K^5} \left(1+5\frac{k_3+k_4}{K} +30 \frac{k_3 k_4}{K^2}\right)+ 23\, {\rm perms.}  \label{Tc4}  \\
T_{c5}&=& (\bkone \cdot \bktwo )( (\bkthree \cdot \bkfour)^2-k_3^2 k_4^2) \times \nn \\
& &\frac{1}{K^3}  \left(1+3 \frac{\sum_{i< j} k_i k_j}{K^2}+15 \left(\sum_i \frac{1}{k_i}\right) \frac{\prod_j k_j}{K^3}+90 \frac{\prod_i k_i}{K^4} \right)+ 23\, {\rm perms.}  \label{Tc5}
\eea
for the trispectra generated respectively by the higher-derivative operators ${\dot \pi}^2((\partial_i \partial_j \p)^2-(\partial_i^2 \p)^2)$ and $(\partial_k \pi)^2((\partial_i \partial_j \p)^2-(\partial_i^2 \p)^2)$ (here $K=k_1+k_2+k_3+k_4$ and the `${\rm perms.}$' stands for the permutations of the 4 wavevectors $\bk_i$). We leave the explicit expressions of the dimensionless coefficients $A_{ci}$ and $A_{sj}$ (functions of $\alpha$) to the appendix \ref{Appendix:trispectrum}, where we also give details about the calculations leading to the result \refeq{result-T}. Besides the appearance of the new shapes \refeq{Tc4} and \refeq{Tc5}, let us stress one important difference compared to the trispectrum generated in k-inflation. There, the specific combination of the coefficients of the cubic action and of $((\partial_i \p)^2)^2$ and $\dot \p^2 (\partial_i \pi)^2$ are such that these two operators cancel at leading order in the quartic Hamiltonian. Such cancellations do not arise in our model, where $T_{c2}$ and $T_{c3}$ contribute at leading-order (\textit{i.e.} as $1/\c^4$) to the trispectrum \refeq{result-T}.

\section{Observational constraints}
\label{constraints}

In its original construction, the orthogonal shape of the bispectrum arises as the outcome of the competition between two equilateral-type shapes \cite{Senatore:2009gt} (although in a substantial fraction of parameter space). This implies that generating an orthogonal bispectrum at the level suggested in WMAP data requires in this context a speed of sound of inflaton perturbations $c_s$ as low as about $0.01$. Unless inflation occurred in the region of parameter space where similar competitions that leads to the orthogonal bispectrum arise at the level of the trispectrum, the amplitude of the latter, scaling as $1/c_s^4$, is of order $10^8$ and hence is comparable to (and greater than) current observational constraint $t_{NL}^{\rm eq}=(-3.11 \pm 7.5) \times 10^6$ \cite{Fergusson:2010gn}. DBI Galileon inflation does not belong to the simple of class of models in which we quantitatively developed this argument in \cite{Renaux-Petel:2013wya}, and the results derived there can not be simply applied here. Yet, we have seen that the two frameworks share similar qualitative features and it is therefore reasonable to expect similar constraints on this model from the trispectrum. We investigate this question in what follows, equipped with the predictions of the preceding section.

\subsection{Constraints from the primordial bispectrum}
\label{constraints-3}

\begin{figure}[!h]
  \center
  \includegraphics[width=0.7\textwidth]{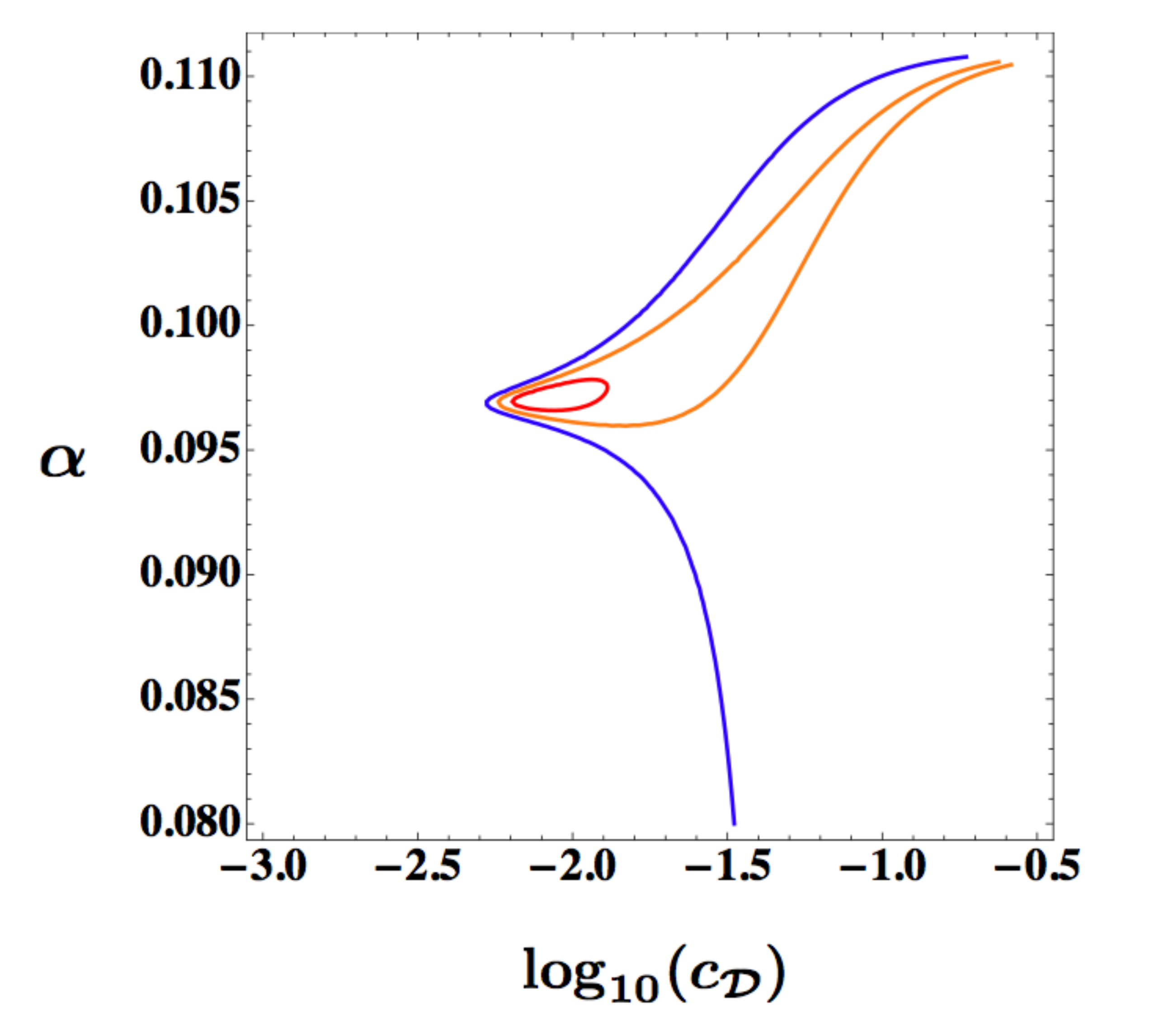}
      \caption{1$\sigma$ (red), 2$\sigma$ (orange) and $3\sigma$ (blue) confidence regions in the $(c_{\cal D},\alpha)$ plane derived from the WMAP9 bispectrum constraints \cite{Bennett:2012fp}.}
      \label{fig:cd-alpha}
\end{figure}

In \cite{Bennett:2012fp}, the WMAP team translated observational constraints on the primordial bispectrum to constraints on the two parameters specifying the cubic action of the simplest effective field theory of inflation. As the bispectrum \refeq{S} generated in DBI Galileon inflation is similar to the one generated in this framework (it is a linear combination of the same two shapes), it is trivial to perform the same analysis here. In particular, Eq.~(57) in \cite{Bennett:2012fp} directly translates here to
\bea
f_{NL}^{{\rm eq}}&=&\frac{1}{\c^2 (1-9 \alpha)}  \left(-0.276\,  \l^2 \,A_{{\dot \pi} (\partial_i \p)^2} -0.0785 \,A_{{\dot \pi}^3}  \right) \label{fNL1} \\
f_{NL}^{{\rm orth}}&=&\frac{1}{\c^2 (1-9 \alpha)}  \left(0.0157 \, \l^2\, A_{{\dot \pi} (\partial_i \p)^2} +0.0163 \,A_{{\dot \pi}^3} \right)\,, \label{fNL2}
\eea
where $A_{{\dot \pi}^3}$, $A_{{\dot \pi} (\partial_i \p)^2}$ and $\lambda$ are given in Eqs.~\refeq{A2}-\refeq{lambda}. This provides the link between the two dimensionless parameters $(\c,\alpha)$ of our model and the two parameters $(f_{NL}^{{\rm eq}},f_{NL}^{{\rm orth}})$ as constrained by WMAP9. Following exactly the same logic as the one in \cite{Bennett:2012fp}, we then easily deduce the $1\sigma$, $2\sigma$ and $3\sigma$ confidence regions displayed in Fig.~\ref{fig:cd-alpha}.
They are qualitatively similar to the constraints on the parameters of the simplest effective field theory of inflation $(c_s,A)$ given in \cite{Bennett:2012fp}. In particular, the $1\sigma$ region favored by data (in red) corresponds to low values of $c_{{\cal D}}$ of about 0.01 and to the narrow region $0.096 \lesssim \alpha \lesssim 0.098$ where the bispectrum \refeq{S} significantly overlaps with the orthogonal template (see Fig.~\ref{fig:Bispectrum-correlations}). The fact that a negative $f_{NL}^{{\rm orth}}$ is favored at $2\sigma$ translates at this level of confidence into the lower bound $\alpha \gtrsim 0.095$ (in orange). Note also that as $\alpha \to 1/9$, the amplitude of $f_{NL}^{{\rm eq}}$ and  $f_{NL}^{{\rm orth}}$ in Eqs.~\refeq{fNL1}-\refeq{fNL2} grow unboundedly, corresponding to a strongly coupled region. This explains the sharpness of the bound on $c_{\cal D}$ in the upper half region. Conversely, for $\alpha \lesssim 0.090$, the bispectrum \refeq{S} is of equilateral type and is nearly insensitive to the strength of the induced gravity.

\subsection{Constraints from the primordial trispectrum}
\label{constraints-4}

\begin{figure}[!h]
\centering
\subfigure[$ | t_{NL}-t_{NL}^{\star} | /\Delta t_{NL}$ in the $3\sigma$ region allowed by the bispectrum constraints, for $\alpha> 0.095$.]{
\includegraphics[width=0.7\textwidth]{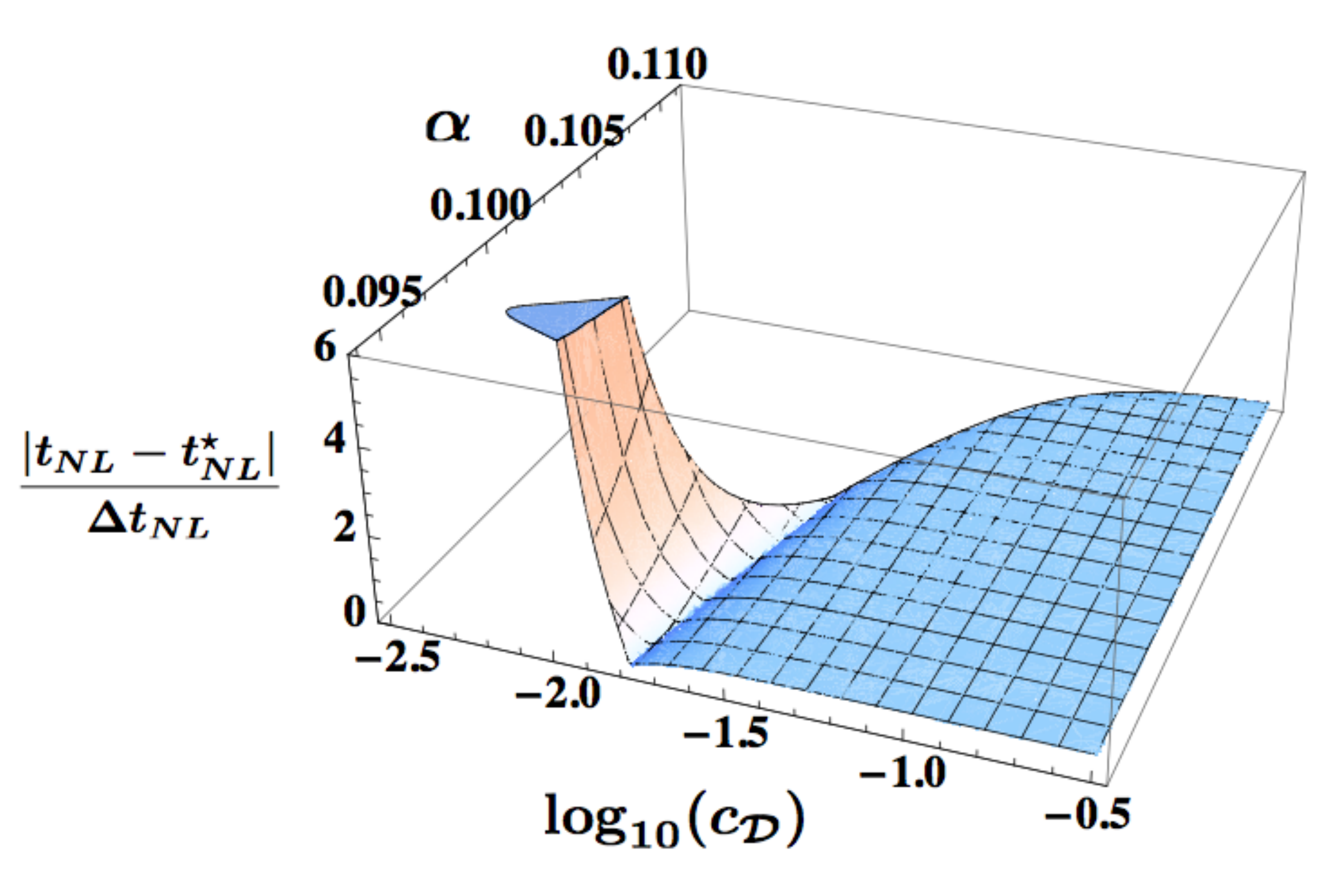}
\label{fig:subfig1}
}
\subfigure[$ | t_{NL}-t_{NL}^{\star} | /\Delta t_{NL}$ in the $1\sigma$ region allowed by the bispectrum constraints.]{
\includegraphics[width=0.7\textwidth]{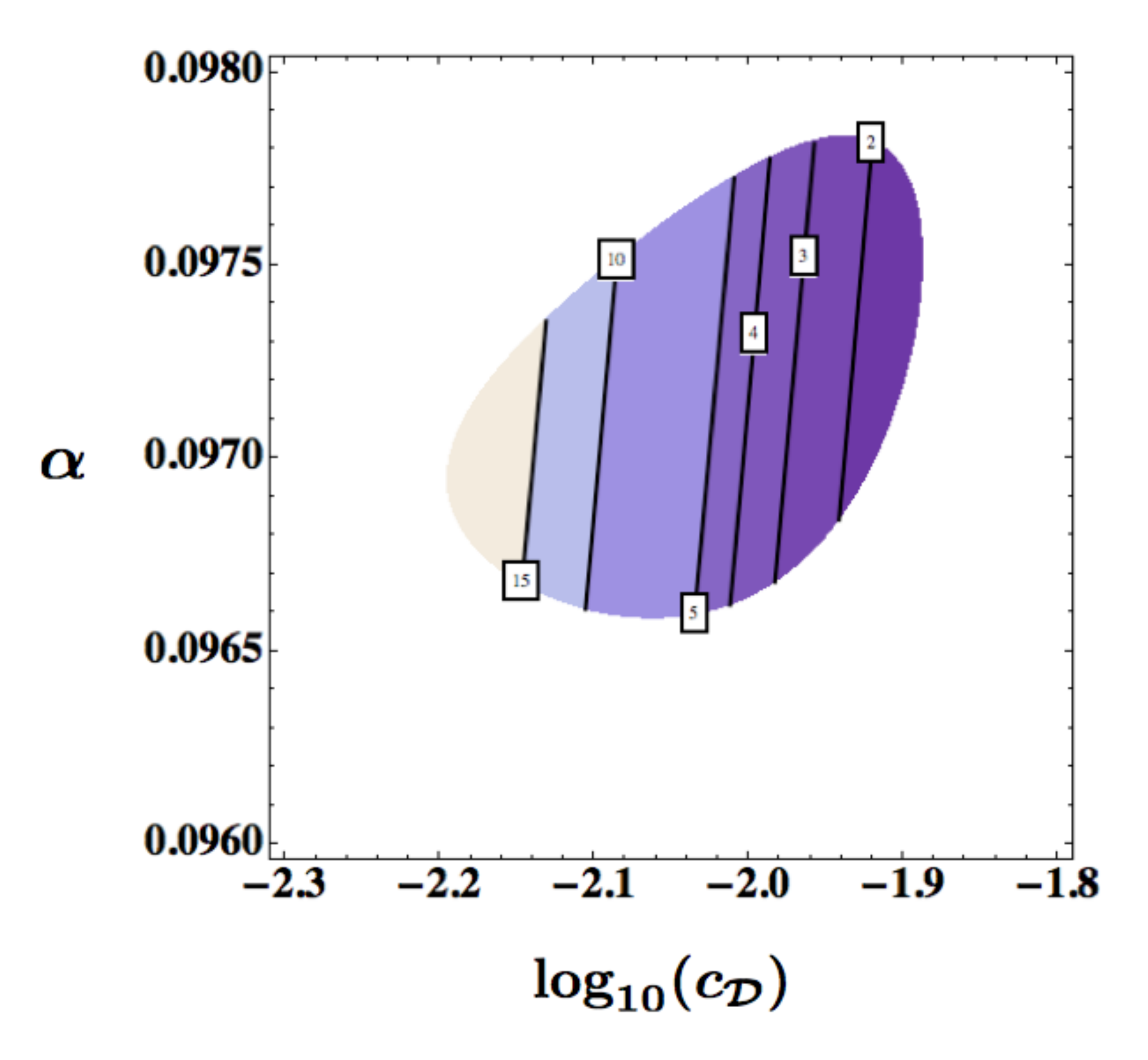}
\label{fig:subfig2}
}
\caption[Optional caption for list of figures]{In the plane $(\c,\alpha)$, confidence level $ | t_{NL}-t_{NL}^{\star} | /\Delta t_{NL}$ at which DBI Galileon inflation is excluded (or not) from the observational constraint on the trispectrum.}
\label{fig:Trispectrum}
\end{figure}

A simple measure of the amplitude of the trispectrum is given by the 
parameter $t_{NL}$ defined such that \cite{Chen:2009bc}
\bea
\frac{1}{k^3} {\cal T}(k_1,k_2,k_3,k_4,k_{12},k_{14}) \xrightarrow[\rm limit]{\rm RT} t_{NL}\,,
\label{tNL}
\eea
where ${\rm RT}$ stands for the regular tetrahedron limit $k_1=k_2=k_3=k_4=k_{12}=k_{14}\equiv k$. While the evaluation at this single point might seem rather ad-hoc, it can be motivated by the fact that trispectra of quantum origin generated around the time of Hubble crossing are generally the largest for the configurations where both the external and internal momenta are of similar magnitude, \textit{i.e.} near this regular tetrahedron limit. This was shown explicitly in reference \cite{Chen:2009bc} for the trispectra $T_{c1}, T_{c2}, T_{c3}, T_{s1}, T_{s2}$ and $T_{s3}$, and this can easily be checked to hold as well for $T_{c4}$ and $T_{c5}$ in Eqs.~\refeq{Tc4}-\refeq{Tc5}.

Using this similarity, the observational constraint relevant for our purpose, $t_{NL}^{\rm eq}=(-3.11 \pm 7.5) \times 10^6$, was derived in \cite{Fergusson:2010gn} (based on methods developed in \cite{Regan:2010cn}) under the assumption that the trispectrum shape ${\cal T}$ in Eq.~\refeq{T-def} can be well approximated by the simple and representative `equilateral-type' shape $T_{c1}$ generated by the operator ${\dot \pi}^4$:
\be
T_{c1}=36 \frac{(k_1 k_2 k_3 k_4)^2}{(k_1+k_2+k_3+k_4)^5}\,.
\label{Tc1}
\ee
More precisely, this constraint -- the only one to date on a trispectrum from quantum origin -- puts a bound on the parameter $t_{NL}^{{\rm eq}}$ defined such that ${\cal T}=t_{NL}^{\rm eq}/ t_{NL}(T_{c1}) \times T_{c1}$. It is therefore only applicable in our model for values of $\alpha$ such that the trispectrum \refeq{result-T} can be well approximated by $T_{c1}$. Fortunately, we have checked this to be the case in the interesting region $\alpha \gtrsim 0.095$ favored (at $2\sigma$) by the observational constraints on the bispectrum (see Fig.~\ref{fig:cd-alpha} in the preceding subsection)\footnote{Note also that for $\alpha \lesssim 0.095$, allowed values of $\c$ from bispectrum constraints are such that the amplitude of the trispectrum is too low to be constrained by data.}. For this purpose, we used the same strategy as in \cite{Renaux-Petel:2013wya} and represented 2-dimensional slices of our trispectrum \refeq{result-T} (as a function of $\alpha$) in different representative limits, as was originally done in \cite{Chen:2009bc}. We display the outcome of this procedure for the representative value $\alpha=0.097$ (in the middle of the $1\sigma$ region favored by the bispectrum) in appendix \ref{Appendix:plots}, from which the approximate proportionality with $T_{c1}$ is manifest.

To summarize, the possibility -- highlighted in a different context in \cite{Renaux-Petel:2013wya} -- of partial cancellations leading to the emergence of an `orthogonal-type' trispectrum qualitatively different from $T_{c1}$ does not arise in DBI Galileon inflation in the interesting range $\alpha \gtrsim 0.095$. The estimator \refeq{tNL} can hence be reliably compared in this regime to the observational bound $t_{NL}^{\rm eq}=(-3.11 \pm 7.5) \times 10^6$ (68 \%  CL). We show in Fig.~\ref{fig:Trispectrum} the confidence level at which DBI Galileon is excluded (or not) by this trispectrum constraint, \textit{i.e.} the represented quantity is $ | t_{NL}-t_{NL}^{\star} | /\Delta t_{NL}$, where $t_{NL}^{\star}=-3.11 \times 10^6$ and $\Delta t_{NL}=7.5 \times 10^6$. The upper figure (respectively the lower figure) displays the whole $3\sigma$ region (respectively zooms in the $1\sigma$ region) favored by the bispectrum constraints, for $\alpha> 0.095$. The conclusion is clear: within the $1\sigma$ region favored by observational constraints on the bispectrum, most of the parameter space $(\c,\alpha)$ is already excluded by the observational bound on the trispectrum, while the amplitude of the latter is too small outside this region to be efficiently constrained by data.

\section{Conclusions}
\label{sec:conclusion}

The DBI Galileon model, in which an induced gravity term is added to the DBI action, is one of the few known explicit inflationary scenarios able to generate a large primordial bispectrum with a significant overlap with the orthogonal template. As the latter was first motivated in the formalism of the effective field theory of inflation, we formulated DBI Galileon in this language. In particular, we identified the various operators and related mass scales that specify the behavior of its fluctuations in the unitary gauge. As a consequence of the presence of higher-derivative operators, we have seen that the way the orthogonal shape arises here differs in details from the one in its original construction. Furthermore, we determined the leading-order action quartic in fluctuations, and calculated the corresponding trispectrum. In the region of parameter space in which the orthogonal-type bispectrum is generated, we found that the former can be very well approximated by the simple `equilateral' trispectrum used in current data analysis. Translating the corresponding observational bound on constraints on the parameters of the model, we showed the combined consideration of the bi- and trispectrum to be very discriminating. In particular, it is only in a small window of parameter space that DBI Galileon inflation can both generate a bispectrum within the $1\sigma$ region favored by WMAP9 data without at the same time generating too large a trispectrum. It will certainly be interesting to compare the predictions derived here to the results of future observations, and in particular to the forthcoming ones of Planck.

On a more theoretical side, we believe several aspects our calculations to be interesting in their own right. In particular, although we considered an inflationary model, we actually made use of tools developed in the context of the effective field theory of dark energy \cite{Gubitosi:2012hu,Bloomfield:2012ff} (see \cite{Creminelli:2006xe} for the first paper in this context). The reason behind this is that some aspects of the effective field theory of inflation, as formulated in \cite{Cheung:2007st}, are useful for classification purposes in a general formalism, but may be less so for practical calculations in concrete models. For instance, although it is always possible, through a conformal transformation of the metric, to set to one the function in front of the Ricci scalar in Eq.~\refeq{S-EFT}, it turned out to be simpler not to do so here,  related to the fact that there is no Einstein frame in the DBI Galileon model. Similarly, although one can always trade the 3d Riemann tensor and covariant derivative for their 4d counterparts, keeping both type of quantities greatly simplified our calculations. Eventually, note that, by using the linear equation of motion to simplify the interacting action, we showed that the presence of higher-derivative operators in the unitary gauge action Eq.~\refeq{Main} manifests itself as higher-derivative $\pi$ interactions at the level of the quartic action only, and it would be interesting to study this in more generality.

\medskip

\begin{acknowledgments}

We would like to thank David Langlois, Kazuya Koyama, Shuntaro Mizuno, Guido W. Pettinari and Filippo Vernizzi for useful conversations related to the topic of this paper, as well as the anonymous referee for helping improving some explanations. This work was supported by French state funds managed by the ANR within the Investissements d'Avenir programme under reference ANR-11-IDEX-0004-02.

\end{acknowledgments}

\appendix

\section{Calculations in the effective field theory formalism}
\label{Appendix:calculations}

In this appendix, we give some details about the calculations performed in subsection \ref{subsec:pi}. \\

\noindent The derivation of Eqs.~\refeq{S-EFT} and \refeq{Main} from the action Eq.~\refeq{S-unitary} follows the steps described in \cite{Cheung:2007st} to show that any single-clock model of inflation can be cast in the form \refeq{S-EFT}. In more details: the way we wrote down Eq.~\refeq{S-unitary} is such that its first line is already in the desired form. As $R^{(3)}$ has no background value, \textit{i.e.} $\delta R^{(3)}=R^{(3)}$, the third line is such that it only contributes to $S^{(n\geq 2)}$, as $\delta R^{(3)}$ times powers of $\delta g^{00}$. The last line, corresponding to the DBI scenario, without induced gravity, can be simply expanded in powers of $\delta g^{00}$. For the second line, we conveniently write 
\bea
K_{ij} K^{ij}-K^2=6 H^2 -4 H K +\delta K_{ij} \delta K^{ij}-(\delta K)^2
\eea
where terms linear in $K$, here and coming from the fourth line, are dealt with using
\bea
\int d^4x \sqrt{-g} \, h(t) K^{\mu}_{\mu}=\int d^4x \sqrt{-g} \, h \nabla_{\mu} n^{\mu}=-\int d^4x \sqrt{-g} \, n^{\mu} \partial_{\mu} h=-\int d^4x \sqrt{-g} \, \sqrt{-g^{00}} {\dot h} \nn\,,
\eea
which is in a form ready to be expanded in power of $\delta g^{00}$. Following this, we easily find Eqs.~\refeq{S-EFT} and \refeq{Main} with
\bea
 \g(t)&=&1+\c \frac{\tM^2}{M_p^2} \label{g} \\
\Lambda(t)&=&V+\frac{1}{2 f \c}\left(1-\c \right)^2\nonumber \\
&+&\frac{\tM^2}{2}\left[{\ddc}+3H {\dc}-2 \Tdot{\left(H\left(\frac{1}{\c}-\c\right) \right)} -\frac{3 H^2}{\c^3}(1+2 \c^2)(1-\c^2) \right] \label{Lambda} \\
c(t)&=&\frac{\c}{2 f}\left(\inv \right)+\frac{\tM^2}{2} \left[-{\ddc}+3H {\dc} +2 \Tdot{\left(H\left(\frac{1}{\c}-\c\right) \right)}  -\frac{3H^2(1-\c^2)}{\c^3}  \right] \label{c}
\eea
where we made no approximation. As explained in \cite{Gubitosi:2012hu} (see also \cite{Bloomfield:2012ff} for an equivalent discussion), the gravitational equations of motions derived from an action of the type \refeq{S-EFT} reads, in a spatially flat FLRW background:
\bea
c&=&\M^2 \, \g \left( -\dot H-\frac12 \frac{\ddot \g}{\g}+\frac{H}{2} \frac{\dot \g}{\g}  \right)\,, \\
\Lambda&=&\M^2 \, \g \left( \dot H+3 H^2+\frac12 \frac{\ddot \g}{\g}+ \frac{5H}{2} \frac{\dot \g}{\g} \right)\,.
\eea
We have verified indeed that these two equations, with $g(t)$, $\Lambda(t)$ and $c(t)$ given in Eqs.~\refeq{g}, \refeq{Lambda} and \refeq{c} respectively, are equivalent to the ones given in \refeq{Friedmann1}-\refeq{Hdot}. This provides a good consistency check of our calculations.

\section{Calculations of the quartic action and of the trispectrum}
\label{Appendix:trispectrum}

In this appendix, we give some details about the calculations of the quartic action and of the trispectrum. \\

\noindent \textbullet\, {\bf Simplification of the quartic action:}\\

\noindent Between Eq.~\refeq{S4-1} and Eq.~\refeq{S4-2}, we traded the operators in ${\dot \pi}^3 \partial_i^2 \pi$ and  $\dot \p (\partial_i \p)^2 \partial_j^2 \p$ for well known ones generated in k-inflation. For this purpose, we used the (leading-order) linear equation of motion \refeq{linear-eom} to replace instances of $\partial_i^2 \pi/a^2$ by $ \Tdot{(a^3 {\dot \pi})}/(a^3 c_s^2)$. Integrating by parts in time, one then finds
\bea
\int dt \, d^3 x \, a^3\,  {\dot \pi}^3 \frac{\partial_i^2 \pi}{a^2}&=&  \int dt \, d^3 x \, a^3\,  \frac{9H}{4 c_s^2} {\dot \pi}^4 \,.
\eea
Using this result, together with 
\be
\int dt \, d^3 x \,a^3\, {\dot \pi}^2  \frac{\partial^i \pi \partial_i {\dot \pi}}{a^2}=-\frac13 \int dt \, d^3 x \,a^3\, {\dot \pi}^3 \frac{\partial_i^2 \pi}{a^2}\,,
\ee
obtained by simple spatial integrations by part, one similarly finds
\bea
\int dt \, d^3 x \, a^3\, \dot \p \frac{(\partial_i \p)^2 \partial_j^2 \p}{a^4}&=&  \int dt \, d^3 x \, a^3\, \left(  \frac{5H}{2c_s^2} {\dot \pi}^2 \frac{(\partial_i \pi)^2}{a^2}  +\frac{3H}{4 c_s^4}{\dot \pi}^4 \right) \,.
\eea

\noindent \textbullet\, {\bf Calculation of the trispectrum:}\\

\noindent Once having determined the second-, third-, and fourth-order action of cosmological perturbations, here in Eqs.~\refeq{S2-1}, \refeq{S3-2} and \refeq{S4-2} respectively, the procedure for calculating the primordial trispectrum is well known. It is based on the In-In (also known as Keldysh-Schwinger) formalism \cite{Weinberg:2005vy}, according to which the expectation value of any operator $Q(t)$ in the interacting vacuum of the theory reads
\be
\langle Q(t) \rangle = \langle 0| \left[ \bar T \exp \left( i \int_{-\infty(1+i \epsilon)}^t H_I(t') dt' \right) \right]
~Q^I(t)~
\left[ T \exp \left( -i \int_{-\infty(1-i \epsilon)}^t H_I(t'') dt'' \right)\right] |0\rangle \,, \label{in-in}
\ee
where $ |0\rangle$ denotes the vacuum of the free theory, $T$ the time-ordering product, $H_I$ the interacting Hamiltonian, $Q^I(t)$ is in the interacting picture, and the $i \epsilon$ indicates that the time integration contour should be slightly rotated into the imaginary plane in order to project onto the interacting vacuum \cite{Maldacena:2002vr}. While the cubic Hamiltonian $H_{(3)}$ is simply $-L_{(3)}$, this simple relation does not hold at quartic order. Instead, a generic Lagrangian of the form
\bea
\CL_2 &=& f_0 \dot \pi^2 + j_2 ~,
\\
\CL_3 &=& g_0 \dot \pi^3 + g_1 \dot \pi^2 + g_2 \dot \pi + j_3
~,
\\
\CL_4 &=& h_0 \dot \pi^4 + h_1 \dot \pi^3 + h_2 \dot \pi^2 + h_3
\dot \zeta + j_4 ~
\eea
gives the fourth-order interaction Hamiltonian density \cite{Huang:2006eha}:
\bea
\CH_4^{I} &=& \left( \frac{9g_0^2}{4f_0} - h_0 \right) \dot\pi_I^4
+ \left( \frac{3g_0g_1}{f_0} - h_1 \right) \dot \pi_I^3
\cr
&+& \left( \frac{3g_0g_2}{2f_0} + \frac{g_1^2}{f_0} -
h_2 \right) \dot\pi_I^2
+ \left(\frac{g_1g_2}{f_0} - h_3 \right) \dot \pi_I
+ \frac{g_2^2}{4f_0} - j_4 ~,
\eea
where $f$, $g$, $h$ and $j$'s are functions of $\pi$, $\partial_i
\pi$ and $t$, and the subscripts denote the orders of $\pi$. In our context, this gives
\bea
\CH_4^{I}&=& \frac{a^3 \epsilon H^2 \M^2}{4 c_s^2} \frac{1-5 \alpha}{1-3\alpha} \frac{1}{\c^4}\left[-4 A_{c1}\, \dot \pi_I^4
+4 A_{c2} \,c_s^2  \, \dot \pi_I^2 \frac{(\partial_i \pi_I)^2}{a^2}-A_{c3} \,c_s^4 \,\frac{((\partial_i \pi_I)^2)^2}{a^4} 
\right.
 \cr
  &&   + 
   \left.
\frac{1}{3} \frac{c_s^4}{H^2} \left( A_{c4}\,{\dot \pi_I}^2 -3 c_s^2 \,A_{c5}\,   \frac{(\partial_i \pi_I)^2}{a^2}  \right) \frac{(\partial_i \partial_j \p)^2-(\partial_i^2 \p)^2}{a^4}
 \right]
 \label{H4}
\eea
where
\bea
A_{c1}&=&\frac{1}{4(1-9\alpha)}\left[5\left(1- \a \left(21-18 \l^2+\frac{18}{5} \l^4  \right)\right)-\frac{9A_{{\dot \pi}^3}^2}{1-9\alpha}   \right] \\
A_{c2}&=&\frac{3}{2(1-5\alpha)} \left[  \left(1-5 \alpha \left(3-2 \l^2 \right)\right)- \frac{A_{{\dot \pi}^3} A_{{\dot \pi} (\partial_i \p)^2}}{1-9\alpha}   \right] \\
A_{c3}&=&\frac{1}{(1-5\alpha)^2} \left[(1-9\alpha)^2-A_{{\dot \pi} (\partial_i \p)^2}^2  \right] \\
A_{c4}&=&-\frac{18 \alpha (1-9\alpha)}{(1-5\alpha)^2} \\
A_{c5}&=&-\frac{2\alpha(1-9\alpha)^2}{(1-5\alpha)^3} \\
A_{s1}&=&\frac{A_{{\dot \pi}^3}^2}{4(1-9\alpha)^2} \\
A_{s2}&=&\frac{A_{{\dot \pi}^3} A_{{\dot \pi} (\partial_i \p)^2}}{2(1-5\alpha)(1-9\alpha)} \\
A_{s3}&=&\frac{A_{{\dot \pi} (\partial_i \p)^2}^2}{(1-5\alpha)^2 }
\eea
and the expressions of $A_{{\dot \pi}^3}$ and $A_{{\dot \pi} (\partial_i \p)^2}$ are given in Eqs.~\refeq{A2}-\refeq{A1}. Note that, up to the overall factor $1/\c^4$, which set the amplitude of the trispectrum, we expressed Eq.~\refeq{H4} in terms of $c_s$, and not $\c$, and of $\epsilon$ \refeq{eps-approx}. The practical reason is the following: upon using a redefined variable 
\be
{\tilde \zeta}=\sqrt{\frac{1-5\alpha}{1-3 \alpha}}\, \zeta\,,
\label{zeta-tilde}
\ee
the quadratic action \refeq{S2-1} can be cast in the form
\bea
S^{(2)}&=&\int dt \, d^3 x \, \frac{a^3\, \epsilon \M^2}{c_s^3} \left[  {\dot {\tilde \zeta}}^2   -c_s^2  \frac{(\partial_i \tilde \zeta)^2}{a^2} \right]\,,
\label{S2-tilde}
\eea
which is formally equivalent to the quadratic action in terms of $\zeta$ in k-inflation \cite{Garriga:1999vw}. Results about the trispectrum derived in this context \cite{Chen:2009bc} can thus be directly translated here provided we use the variables $\tilde \zeta$ \refeq{zeta-tilde}, $\epsilon$ \refeq{eps-approx} and $c_s$ \refeq{cs} in intermediate calculations. Using the form
\bea
 \hspace{-2.0em}  S^{(3)}_{\rm eff}&=&\int dt \, d^3 x \, \frac{a^3 \epsilon H^2 \M^2}{c_s^2}\,  \frac{1-5 \alpha}{1-3 \alpha} \frac{1}{\c^2 (1-9\alpha)}   \left[A_{{\dot \pi}^3}  \dot \p^3 - \lambda^2 c_s^2 A_{{\dot \pi} (\partial_i \p)^2}  \dot \p \frac{(\partial_i \p)^2}{a^2}    \right]
\eea
of the cubic action Eq.~\refeq{S3-2}, together with Eqs.~\refeq{H4} and \refeq{S2-tilde}, we thus straightforwardly deduce from \cite{Chen:2009bc} the final result Eq.~\refeq{result-T}, except for the contact-interaction trispectra Eqs.~\refeq{Tc4}-\refeq{Tc5} -- coming from the higher-derivative operators ${\dot \pi}^2((\partial_i \partial_j \p)^2-(\partial_i^2 \p)^2)$ and $(\partial_k \pi)^2((\partial_i \partial_j \p)^2-(\partial_i^2 \p)^2)$ absent in k-inflation -- that we calculated from first principles following Eq.~\refeq{in-in}.

\section{Shapes of trispectra}
\label{Appendix:plots}

In this appendix, we represent the trispectrum \refeq{result-T} for the interesting value $\alpha=0.097$ lying in the middle of the $1\sigma$ region favored by observational constraints on the bispectrum (see Fig.~\ref{fig:cd-alpha}), and at which the correlation of the bispectrum with the orthogonal template is the largest (see Fig.~\ref{fig:Bispectrum-correlations}). We set $\c=1$ without losing generality and follow the same strategy as in \cite{Renaux-Petel:2013wya}: we represent 2-dimensional slices of relevant trispectra in different representative limits, as was originally done in \cite{Chen:2009bc}. Note also that we chose to plot the scale-independent quantities ${\cal {\tilde T}}={\cal T}/(k_1 k_2 k_3 k_4)^{3/4}$ rather than ${\cal T}$ itself. The four limits we considered are:\\

  \textbullet \,\, \ The specialized planar limit, in which $k_1=k_3=k_{14}$, and the
    tetrahedron reduces to a planar quadrangle with \cite{Chen:2009bc} 
\begin{align}
      k_{12}=\left[
k_1^2+\frac{k_2 k_4}{2 k_1^2}\left( k_2 k_4 +
\sqrt{(4k_1^2-k_2^2)(4k_1^2-k_4^2)} \right) \right]^{1/2}~.
    \end{align}
   Shapes are then represented as functions of $k_2/k_1$ and $k_4/k_1$.

  \textbullet \,\, \  Near the double-squeezed limit: ${k}_3={k}_4=k_{12}$ and the tetrahedron
  is a planar quadrangle with \cite{Chen:2009bc} 
\begin{align}\label{planark2}
   k_2= \frac{\sqrt{k_1^2 \left(-k_{12}^2+k_3^2+k_4^2\right)- k_{s1}^2 k_{s2}^2+k_{12}^2 k_{14}^2+k_{12}^2 k_4^2+k_{14}^2
   k_4^2-k_{14}^2 k_3^2-k_4^4+k_3^2 k_4^2}}{\sqrt{2} k_4}~,
  \end{align}
where $k_{s1}$ and $k_{s2}$ are defined as
\begin{align}
&  k_{s1}^2\equiv 2\sqrt{(k_1 k_4+{\bf k}_1 \cdot {\bf k}_4)(k_1
k_4-{\bf k}_1 \cdot {\bf
  k}_4)}~,\nonumber\\ &
k_{s2}^2\equiv 2\sqrt{(k_3 k_4+{\bf k}_3 \cdot {\bf k}_4)(k_3
k_4-{\bf k}_3 \cdot {\bf
  k}_4)}~.
\end{align}
 Shapes are then represented as functions of $k_{4}/k_1$ and $k_{14}/k_1$.

  \textbullet \,\, \  The folded limit: $k_{12}=0$, hence $k_1=k_2$ and $k_3=k_4$. Shapes are then represented as functions of $k_4/k_1$ and $k_{14}/k_1$, and we assumed $k_4 < k_1$ without loss of generality.

 \textbullet \,\, \  The equilateral limit: $k_1=k_2=k_3=k_4$. Shapes are then represented as functions of $k_{12}/k_1$ and $k_{14}/k_1$.\\

\begin{figure}[!h]
  \center
  \includegraphics[width=1.0\textwidth]{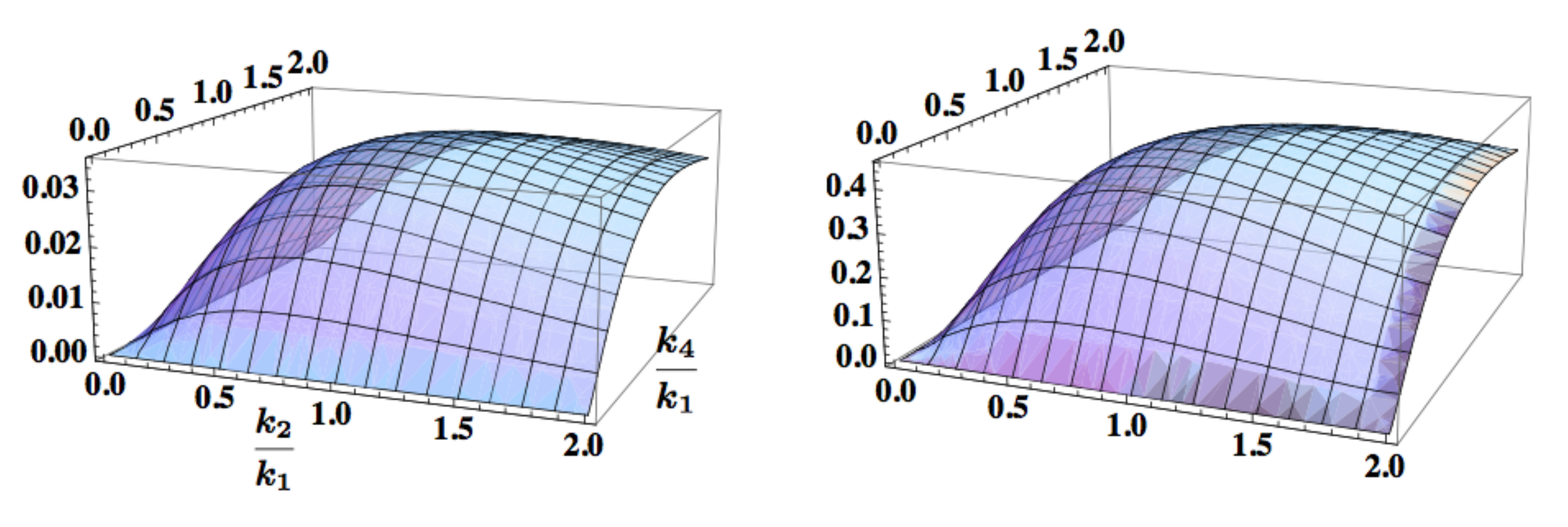}
      \caption{\label{SPL}
         We consider here
      the specialized planar limit with $k_1=k_3=k_{14}$, and plot ${\tilde T}_{c1}$ (left) and ${\cal {\tilde T}}$ in Eq.~\refeq{result-T} for $\alpha=0.097$ (right), as
      functions of $k_{2}/k_1$ and
    $k_{4}/k_1$.}
\end{figure}

\begin{figure}[!h]
  \center
  \includegraphics[width=1.0\textwidth]{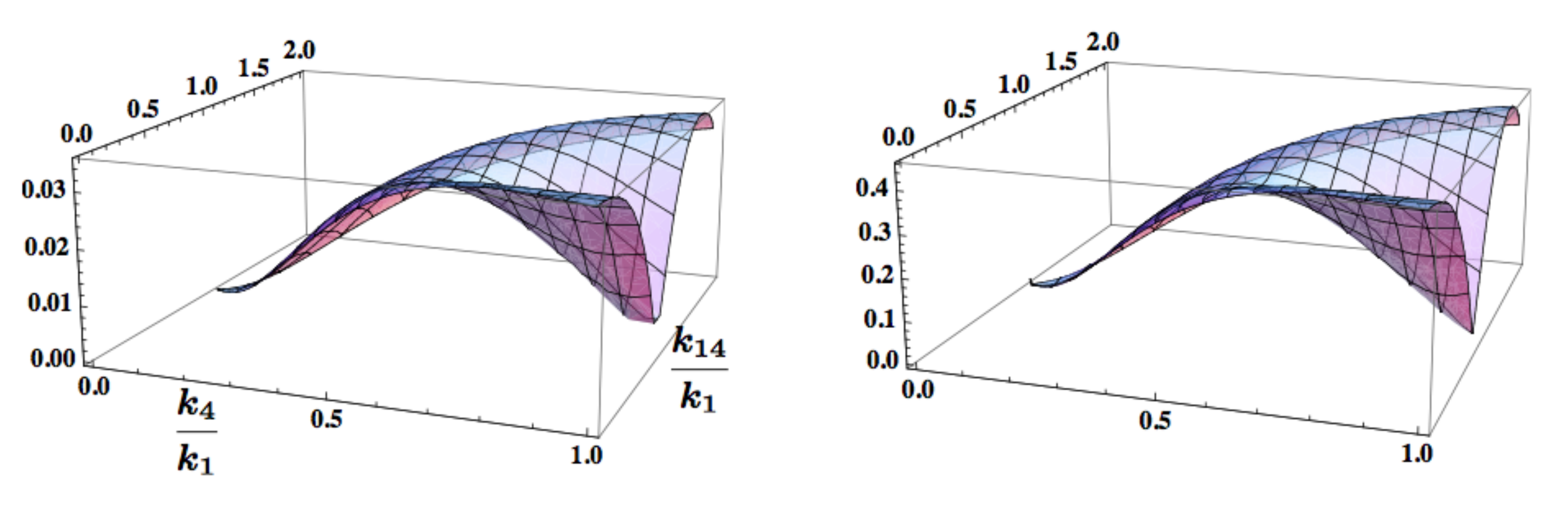}
      \caption{\label{DSL} We look here
      at the shapes near
      the double squeezed limit: we consider the case where
      ${k}_3={k}_4=k_{12}$ and
      the tetrahedron
  is a planar quadrangle. We plot ${\tilde T}_{c1}$ (left) and ${\cal {\tilde T}}$ in Eq.~\refeq{result-T} for $\alpha=0.097$ (right), as
      functions of $k_{4}/k_1$ and $k_{14}/k_1$.}
\end{figure}

\begin{figure}[!h]
  \center
  \includegraphics[width=1.0\textwidth]{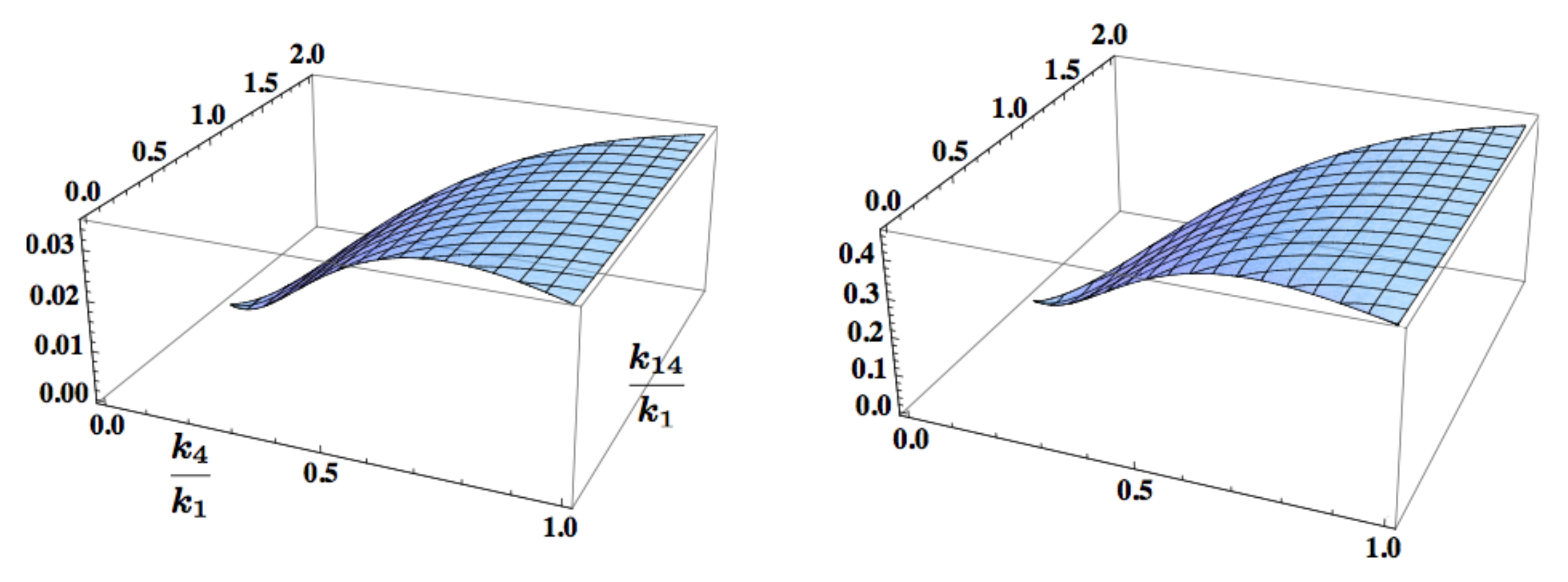}
      \caption{\label{Folded} We consider here
      the folded limit
      $k_{12}=0$, and plot ${\tilde T}_{c1}$ (left) and ${\cal {\tilde T}}$ in Eq.~\refeq{result-T} for $\alpha=0.097$ (right), as
      functions of $k_{4}/k_1$ and
      $k_{14}/k_1$.}
\end{figure}

\begin{figure}[!h]
  \center
  \includegraphics[width=1.0\textwidth]{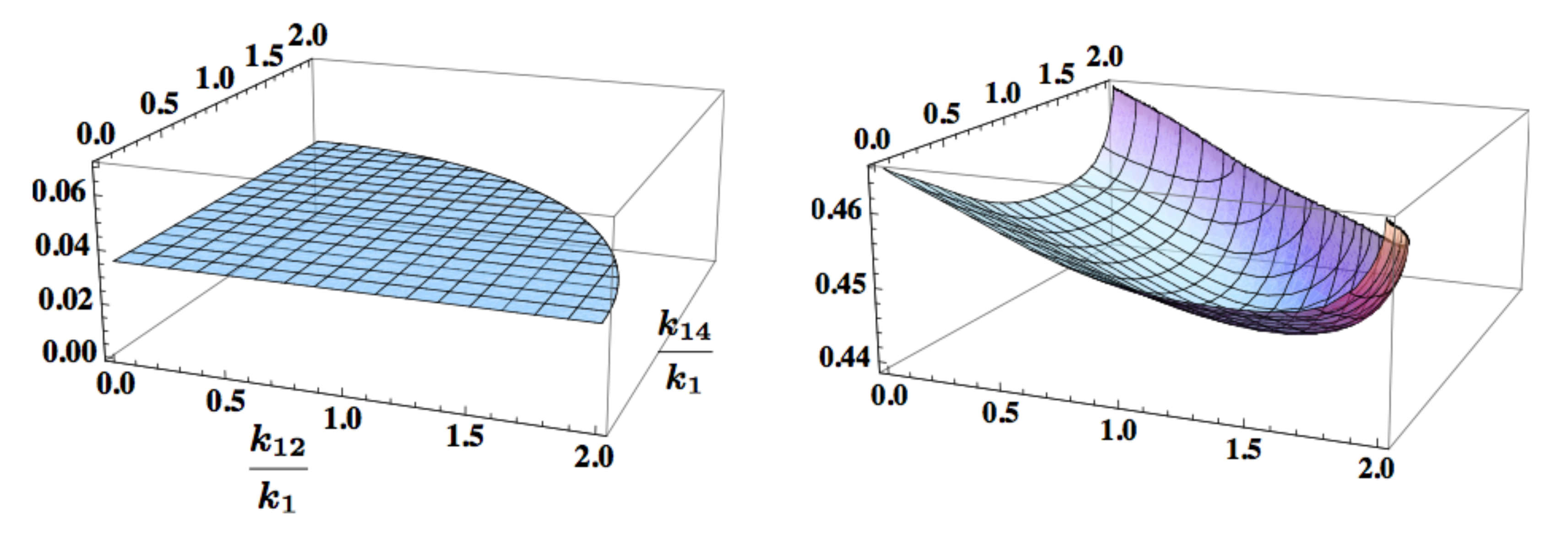}
      \caption{\label{Equilateral} We consider here
      the equilateral limit $k_1=k_2=k_3=k_4$, and plot ${\tilde T}_{c1}$ (left) and ${\cal {\tilde T}}$ in Eq.~\refeq{result-T} for $\alpha=0.097$ (right), as
      functions of $k_{12}/k_1$ and
      $k_{14}/k_1$.}
\end{figure}

In Figs.~\ref{SPL}, \ref{DSL}, \ref{Folded} and \ref{Equilateral}, we represent the (rescaled) shape function ${\cal {\tilde T}}$ in Eq.~\refeq{result-T} for $\alpha=0.097$, and for comparison the representative `equilateral-type' shape ${\tilde T}_{c1}$ currently used in data analysis \cite{Fergusson:2010gn}, in the specialized planar limit, near the double-squeezed limit, in the folded limit and in the equilateral limit respectively. From these plots, the (approximate) proportionality between the two trispectra is obvious. More generally, we have checked that $T_{c1}$ is a good ansatz for the DBI Galileon trispectrum \refeq{result-T} in the whole interesting range $0.095< \alpha< 1/9$.

\end{document}